\documentstyle[prl,aps,multicol,amssymb,epsfig]{revtex}
\def\footnoterule{\kern-3pt \hrule width\hsize \kern3pt}
\tighten
\title{Generation and degree of entanglement in a
relativistic formulation}
\author{Jiannis Pachos\footnotemark[1]$^1$ and Enrique Solano$^{1,2}$}
\address{
$^1$Max-Planck-Institut f\"ur Quantenoptik, Hans-Kopfermann-Strasse 1,
D-85748 Garching,
Germany\\
$^2$Secci\'{o}n F\'{\i}sica, Departamento de Ciencias,
Pontificia Universidad Cat\'{o}lica del Per\'{u}, Apartado 1761,
Lima, Peru
}
\date{\today}

\begin{document}

\newcommand{\PHI}{\mbox{\boldmath${\phi}$}}
\newcommand{\U}{{\cal U}}
\newcommand{\cp}{{\bf CP}}
\newcommand{\be}{\begin{equation}}
\newcommand{\ee}{\end{equation}}
\newcommand{\bq}{\begin{eqnarray}}
\newcommand{\eq}{\end{eqnarray}}
\newcommand{\Sc}{Schr\"odinger\,\,}
\newcommand{\Sp}{\,\,\,\,\,\,}
\newcommand{\no}{\nonumber\\}
\newcommand{\tr}{\text{tr}}
\newcommand{\p}{\partial}
\newcommand{\la}{\lambda}
\newcommand{\La}{\Lambda}
\newcommand{\G}{{\cal G}}
\newcommand{\D}{{\cal D}}
\newcommand{\E}{{\cal E}}
\newcommand{\W}{{\bf W}}
\newcommand{\de}{\delta}
\newcommand{\al}{\alpha}
\newcommand{\bi}{\beta}
\newcommand{\ga}{\gamma}
\newcommand{\ep}{\epsilon}
\newcommand{\si}{\sigma}
\newcommand{\vep}{\varepsilon}
\newcommand{\th}{\theta}
\newcommand{\om}{\omega}
\newcommand{\J}{{\cal J}}
\newcommand{\pr}{\prime}
\newcommand{\ka}{\kappa}
\newcommand{\TH}{\mbox{\boldmath${\theta}$}}
\newcommand{\DE}{\mbox{\boldmath${\delta}$}}
\newcommand{\lan}{\langle}
\newcommand{\ran}{\rangle}
\newcommand{\A}{{\cal A}}

\maketitle
\draft

\thispagestyle{empty}

\begin{abstract}
The generation of entangled states and
their degree of entanglement is studied {\it ab initio}
in a relativistic formulation for the
case of two interacting spin-1/2 charged particles. In the realm
of quantum electrodynamics we derive the interaction that produces
entanglement between the spin components of covariant Dirac
spinors describing the two particles. Following this consistent
approach the relativistic invariance of the generated
entanglement is discussed.
\end{abstract}

\footnotetext[1]{Jiannis.Pachos@mpq.mpg.de}

\vspace*{0.2cm}
\noindent
\pacs{PACS: 03.65.Ud, 03.30.+p, 11.30.Cp}

\begin{multicols}{2}

The generation of entangled states and the measurement of their
degree of entanglement are at the heart of diverse fundamental
tests of quantum mechanics. On the other hand, special relativity
is a fundamental theory that has to be considered in the study of
measurements realized by different moving observers. In special
relativity, for example, the observed simultaneity of two space-like
separated events can be broken when they are observed in a
different reference frame. As a consequence, there is a natural
interest in studying nonlocal quantum correlations in the
framework of special relativity~\cite{czachor}.
Recent experiments~\cite{gisin} have addressed
the question about the compatibility of the apparently independent
predictions of quantum mechanics and special relativity. In
Ref.~\cite{peres}, Peres {\it et al.} have shown that the spin
entropy of a single free spin-$\frac{1}{2}$ particle has no
invariant meaning when the kinematical degrees of freedom are traced
out. More recently, in Ref.~\cite{gingrich}, Gingrich and
Adami have studied the transfer of entanglement between momentum
and spin of two particles under a Lorentz transformation.

In this paper a consistent approach is presented to the generation
and measurement of entanglement in a quantum relativistic framework.
As an illustrative example, the case of two interacting spin-1/2
massive particles, say electrons, is considered in the context of
quantum electrodynamics (QED). We derive the spin interaction that
produces entanglement, maximal or not, between the spin components of
the covariant Dirac spinors associated with the two particles. 
The main goal is to keep throughout the process a relativistic
covariant formalism for the electrons and to obtain the
quantum interactions that generate entanglement between them. This
{\it ab initio} procedure will permit us to discuss consistently the
invariant properties of the degree of entanglement of the generated
states~\cite{milburn}.

Let us consider two non-relativistic spin-1/2 charged particles,
e.g. electrons,
$1$ and $2$ with charge $e$ and spin states given in the basis of the
$z$ axis eigenstates $| \!\! \uparrow \rangle=(1,0)^T$ and $| \!\!
\downarrow \rangle=(0,1)^T$. For the system of the two-particle spin
states we
construct the spin tensor product states $| i j \rangle = | i \rangle
\otimes | j \rangle$ with $ i,j = \{ \uparrow, \downarrow \}$. The
magnetic dipole-dipole interaction between the spins of two particles,
in nonrelativistic quantum mechanics, is described by the Hamiltonian
\begin{eqnarray}
H&&=-\text{\mbox{\boldmath$\mu$}}_1 \cdot
\text{\mbox{\boldmath$\nabla$}} \times
\Big(\text{\mbox{\boldmath$\mu$}}_2 \times
\text{\mbox{\boldmath$\nabla$}} { 1 \over 4 \pi r}\Big) 
\no \no 
&& ={3 ({\bf n} \cdot\text{\mbox{\boldmath$\mu$}}_1 ) ({\bf n}
\cdot\text{\mbox{\boldmath$\mu$}}_2 )
-\text{\mbox{\boldmath$\mu$}}_1
 \cdot\text{\mbox{\boldmath$\mu$}}_2  \over 4 \pi r^3}  + {2 \over 3}
\text{\mbox{\boldmath$\mu$}}_1
 \cdot\text{\mbox{\boldmath$\mu$}}_2 \,\, \delta ({\bf r})
\label{ham} 
\end{eqnarray}
where $r$ is the distance between the electrons,
${\bf n}$ is a unit vector in the direction of $r$,
\mbox{\boldmath$\mu$}$ = (e \hbar /2 m_e c)$\mbox{\boldmath $
\sigma$} is the dipole moment operator of each electron and
\mbox{\boldmath $ \sigma$} $\equiv {\bf x} \, \sigma ^x +{\bf y} \,
\sigma ^y+{\bf z} \, \sigma ^z$ is the spin operator. 
When Hamiltonian (\ref{ham}) is applied to the initial
state $| \!\! \downarrow \uparrow \rangle$, for non-overlapping
particles and for ${\bf n} ={\bf n}_z$ it generates entanglement 
by means of the term $-\text{\mbox{\boldmath$\mu$}}_1
\cdot\text{\mbox{\boldmath$\mu$}}_2/4 \pi r^3$. 
After an interaction time $t$ an entangled state is produced, which up
to a global phase is given by
\be
|\Psi^-\rangle= \cos(2 J t) | \!\! \downarrow \uparrow
\rangle - i \sin(2 J t) | \!\!  \uparrow \downarrow \rangle
\label{noncov1}
\ee
where the coupling constant $J =-e^2 \hbar^2 /(16 \pi m_e^2 c^2
r^3)$.

The dipole interaction model in nonrelativistic quantum mechanics
involves only the spin degrees of freedom of the two particles and
its compatibility with special relativity, frequently introduced
{\it ad hoc}, should be proven from fundamental principles. In
what follows, we derive a spin interaction Hamiltonian, similar to
Eq.~(1), from first principles and in the realm of quantum
electrodynamics, which describes the interaction in a relativistic
manner. This allows the creation of entangled states, similar to
Eq.~(2), between the spin components of the Dirac spinors of the
two particles, while keeping relativistic covariance in a natural
way.

\begin{center}
\begin{figure}[ht]
\centerline{ \put(30,45){$p_1$} \put(30,120){$p_1^\prime$}
\put(110,45){$p_2$} \put(110,120){$p_2^\prime$}
\put(170,45){$p_1$} \put(170,120){$p_1^\prime$}
\put(250,45){$p_2$} \put(250,120){$p_2^\prime$}
\put(62,70){$p_1^\prime\! - \! p_1$} \put(203,70){$p_1^\prime \! -
\! p_2$} \hspace*{-0.8cm} \epsffile{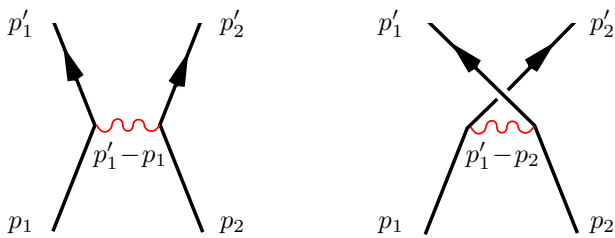} } \vspace{-0.5cm}
\caption[contour]{\label{sca} The scattering process of two particles
by the exchange of one virtual photon. This process incorporates
the dipole-dipole interaction between two spin-1/2 charged particles.}
\end{figure}
\end{center}

In the relativistic quantum domain we consider two identical spin-1/2
charged particles described by Dirac spinors. They may be considered
as relativistic quantum mechanical objects or as one-state particles
of quantum field theory that are allowed to interact with a quantized
electromagnetic field. Their interactions will
be studied within the  covariant formalism of QED by considering
scattering processes. In particular,
the spin interaction Hamiltonian can be derived by calculating the
amplitude that corresponds to Feynman diagrams describing the
scattering of the charged particles when they exchange one virtual
photon \cite{Sakurai}. Two diagrams contribute to this amplitude 
as a consequence of the indistinguishability of the
particles resulting from the fermion statistics (see Fig. \ref{sca}).

\vspace{-1cm}
Each particle participating in the scattering is described by a Dirac
spinor, which for the plane
wave case is given by~\cite{Peskin} \be \psi(x,\epsilon)=
u(p,\epsilon) {\rm e}^{-{\rm i} p \cdot x} , \ee where $x$ is a
point in space-time, $p$ is the energy-momentum of the plane wave
and $\epsilon$ is the polarization of the spin variable. The wave
function $\psi(x,\epsilon)$ satisfies the Dirac equation \be ({\rm
i}\gamma ^\mu \partial_\mu -m)\psi=0 , \ee where $\gamma^\mu$ are
the Dirac matrices. Then, the vector $u(p,\epsilon)$ is a four
dimensional spinor of the form
\[
u(p,\epsilon)=\left( \begin{array}{c} \sqrt{p\cdot \sigma } \,\,
\xi^{\epsilon}
\\ \sqrt{p\cdot \bar
      \sigma } \,\, \xi^{\epsilon} \end{array} \right) ,
\]
where $\sigma^\mu\equiv (\mathbf{1}, \text{\mbox{\boldmath
$\sigma$}})$, $\bar \sigma^\mu\equiv (\mathbf{1}, -
\text{\mbox{\boldmath $\sigma$}})$ and $\xi$ is any
two component spinor normalized to $\xi^\dagger \xi=\mathbf{1}$.
For example, for $\epsilon=\uparrow$, $\xi^\uparrow=(1,0)^T$,
which means that the particle has spin up in the $z$ direction.
The initial state of two well separated particles with known spin
orientation can be expressed as
\begin{equation}
u(p_1,\epsilon_1) \otimes u(p_2,\epsilon_2) .
\end{equation}
In the following, we calculate their scattering amplitude
${\cal{M}}$ without tracing their spin components, by employing
the QED Feynman diagrams and rules. The leading order
contributions to this amplitude in terms of a perturbation
expansion of the QED coupling $e$ are presented in
Fig.~\ref{sca}.
The amplitude of this process is given by
\begin{eqnarray}
&&
i{\cal{M}}=
\no\no
&&
(-i e)^2\Big[ \bar u (p^\prime_1,\epsilon_1^\prime) \gamma^\mu
u(p_1,\epsilon_1) {-i g_{\mu \nu} \over (p_1^\prime - p_1)^2} \bar u
(p^\prime_2,\epsilon^\prime_2) \gamma^\nu u (p_2,\epsilon_2)
\no \no
&&
-\bar u (p_1^\prime,\epsilon_1^\prime) \gamma^\nu u(p_2, \epsilon_2)
{-i g_{\mu \nu}\over (p_1^\prime - p_2)^2} \bar u
(p^\prime_2,\epsilon_2^\prime) \gamma^\mu u (p_1,\epsilon_1)
  \Big].
\label{ampl} 
 \end{eqnarray}
For nonrelativistic velocities we can employ the Born approximation
for the scattering amplitude, given by $\langle p^\prime | i {\cal M}
| p\rangle=-i \tilde V({\bf q}) \delta^{(3)}(E_{{\bf p}^\prime } -
E_{\bf p})$ for ${\bf q}={\bf p}^\prime -{\bf p}$. This does not
violate the relativistic covariance of the description but rather
isolates the dominant contributions for particles moving slowly in a
specific reference frame. By substituting
for low momentum $u(p^\prime,\epsilon^\prime) \gamma^i u(p,
\epsilon) \approx 0$ for $i=1,2,3$ and
$u(p^\prime,\epsilon^\prime) \gamma^0 u(p, \epsilon) \approx 1$ we
obtain by Fourier transformation the usual Coulomb interaction
$
V(r)=(2m)^2 e^2 / (4 \pi r) \delta^{{\epsilon}_1
\epsilon_1^\prime} \delta^{\epsilon_2 \epsilon_2^\prime} . 
$
The Kronecker $\delta$ symbols indicate that the spin indeces of the
spinors remain unaffected by the Coulomb interaction.
Here, only the first diagram has been employed as an
antisymmetrization of the resulting wave function automatically
compensates for the contributions due to the exchange diagram. In
the next order of approximation with respect to small momentum, where
still the Born approximation holds, we employ the exact formula
\begin{eqnarray}
&& u( p^\prime,\epsilon^\prime) \gamma^i u(p, \epsilon)= \no
\no && \left. \xi ^{\epsilon_1^\prime }\right.^\dagger \left[- {i
({\bf p}_1+{\bf p}_1^\prime) \over 2 m}+
{\text{\mbox{\boldmath$\sigma$}}_1 \times ({\bf p}_1-{\bf
p}_1^\prime) \over 2m} \right] \xi^{\epsilon_1} \,\, . 
\end{eqnarray}
Considering in the amplitude (\ref{ampl}) only the cross terms with
\mbox{\boldmath$\sigma$}, we 
obtain by Fourier transformation the dipole-dipole interaction
matrix \be -{e \, \text{\mbox{\boldmath$\sigma$}}_1 \over 2m}
\cdot \text{\mbox{\boldmath$\nabla$}} \times \Big({e \,
\text{\mbox{\boldmath$\sigma$}}_2 \over 2m} \times
\text{\mbox{\boldmath$\nabla$}} { 1 \over 4 \pi r}\Big) . \ee This
term is analogous to Eq.~(\ref{ham}), giving rise to the spin
interaction, but this time in the space of spinors. In particular,
if we have an initial two-particle state with non-overlapping wave
functions and spins oriented oppositely along
the $z$ axis, the effective Born potential becomes \be V=J \left(
2 \delta^{\epsilon_1 \epsilon_2^\prime} \delta^{\epsilon_2
\epsilon_1^\prime}- \delta^{{\epsilon}_1 \epsilon_1^\prime}
\delta^{\epsilon_2 \epsilon_2^\prime} \right) \label{spinors} ,\ee
equivalent to the term $-\text{\mbox{\boldmath$\mu$}}_1
\cdot\text{\mbox{\boldmath$\mu$}}_2/4 \pi r^3$ of Eq.~(\ref{ham})
but acting on four-component Dirac spinors. If the relativistic
description of the particles is relaxed, the spin interaction of
Eq. (\ref{ham}) is recovered. For the initial two-particle state
$u({\bf p}_1=0,\epsilon_1=\downarrow)$ $\otimes u({\bf
p}_2=0,\epsilon_2=\uparrow)$, the evolved state after an
interaction time $t$ reads 
\begin{eqnarray}
&& \cos
(2Jt) \,\, u({\bf p}_1=0,\epsilon_1=\downarrow)\otimes u({\bf
p}_2=0,\epsilon_2=\uparrow) \no \no && \,\,\,\, -i \sin (2J t)
\,\, u({\bf p}_1=0,\epsilon_1=\uparrow)\otimes u({\bf
p}_2=0,\epsilon_2=\downarrow) \,\, , \label{cov1} 
\end{eqnarray}
up to an overall phase. Here, we have assumed that the particles are
asymptotically at rest with respect to a common reference frame.
The relativistic covariance of the entangled two-particle state of
Eq.~(\ref{cov1}) is certainly not present in Eq.~(\ref{noncov1}).
Notice that the form of the coherent superposition of Eqs. (2) and
(10) is preserved and $J t$ is a relativistic invariant quantity due
to the invariance of the action of the electromagnetic
interactions~\cite{comment1}.

The Dirac equation is relativistically covariant, i.e. it has an
invariant form under transformations from one inertial frame to
the other. Let us consider a spin-1/2 particle with momentum $p$
and spin $\epsilon$ described by the wave function
$\psi(p,\epsilon)$. In a transformed reference frame its momentum
is given by $p^\prime=Lp$, where $L$ is a Lorentz boost
represented by a non-unitary matrix. The quantized Dirac states
$|\psi(p,\epsilon)\rangle=\psi(p,\epsilon) {a_{\bf
p}^\epsilon}^\dagger |0\rangle$, with ${a_{\bf
p}^\epsilon}^\dagger$ being the creation operator of an electron and
$|0\rangle$ the vacuum state, transform unitarily, i.e.
$|\psi'(p',\epsilon')\rangle = U ( L ) |\psi(p,\epsilon)\rangle$.

We consider now the initial wave function of the two particles
$u(p_1,\epsilon_1)\otimes u(p_2,\epsilon_2)$, which for simplicity
we shall take to have zero three-momentum~\cite{comment2} and with
spin components oriented oppositely along the $z$ axis, i.e.
$\epsilon_1=\downarrow$ and $\epsilon_2=\uparrow$. Then, each
spinor is given by $u({\bf p}=0,\epsilon=\downarrow \uparrow
)=\sqrt{m}(\xi^{\downarrow \uparrow},\xi^{\downarrow
\uparrow})^T$. According to the previous, when the two
particles are allowed to interact for a certain time $T=\pi/(4J)$,
the spinor ``EPR state" 
\begin{eqnarray}
\Psi({\bf p}_1=0,&&{\bf p}_2=0)= {1
\over \sqrt{2}}\Big[ u(0,\epsilon_1=\downarrow)\otimes
u(0,\epsilon_2=\uparrow) \no \no && -i
u(0,\epsilon_1=\uparrow)\otimes u(0,\epsilon_2=\downarrow)\Big]
\end{eqnarray}
is generated. This state corresponds to a Lorentz frame where
both particles are at rest. For this ``EPR state" the kinematical
degrees of freedom are incorporated in a relativistic formalism,
in contrast to Eq. (\ref{noncov1}). Hence, the wave function
of each of the particles can be transformed to a relativistic frame
along the $x$ direction where it becomes \be u({\bf p}=p_x {\bf
x},\epsilon=\downarrow \uparrow)= \left(
\begin{array}{c}
\sqrt{E-p_x \sigma_x} \, \, \,
\xi^{\downarrow \uparrow}\\
\sqrt{E+p_x \sigma_x} \, \, \, \xi^{\downarrow \uparrow}
\end{array}
\right) \,\, . \ee  Note that in general the kinematics and the
spin degrees of freedom of each particle are not factorizable, so
they have to be incorporated in the theory {\it ab initio} in the
entanglement generation procedure. ´An arbitrary
Lorentz transformation that consists of a boost
and a rotation acting on the ``EPR state" gives the state 
\begin{eqnarray}
\Psi({\bf p}_1={\bf p},&&{\bf p}_2={\bf p}) =
{1 \over \sqrt{2}} \Big[ u({\bf p},\tilde{\epsilon}_1 =
\downarrow)\otimes u({\bf p},\tilde{\epsilon}_2 = \uparrow) \no
\no && -i u({\bf {\bf p}},\tilde{\epsilon}_1 = \uparrow)\otimes
u({\bf p},\tilde{\epsilon}_1 = \downarrow) \Big] , \label{enta}
\end{eqnarray}
where $\tilde{\epsilon}$ indicates spin up or down in the new
spin direction. Here, we are allowed to make this transformation
as the spinors are written in a relativistic covariant formalism.
The spin states of each particle observed in the moving reference
frame are equivalent to a local unitary transformation of the spin
states observed in the rest frame, thus preserving the degree of
entanglement. Transforming the entangled state to a different
reference frame results into a state that can be observed by moving
detectors. Ideal detectors could distinguish the different
orientations of the spin for different momenta of a particle as
the momentum and the spin operators commute. Therefore, in the
considered situation, different from Ref.~\cite{peres}, even
though the spin is not a relativistic invariant quantity, the
degree of entanglement is.

In this paper, we showed that the magnetic dipole-dipole interaction
employed in nonrelativistic quantum mechanics to describe the spin
entanglement between two spin-1/2 charged particles can be
derived from first principles, i.e. quantum electrodynamics. In
this way, the system can be described in a Lorentz covariant formalism
throughout the whole process. Hence, we were able to produce
entanglement between Dirac spinors and to ask consistently about
its relativistic properties. We showed that, as a natural
consequence, the measurement outcomes of any two moving observers
should witness the same degree of entanglement, independent of
their relative motion. In this relativistic context, it is clear
that no superluminal communication or causality violation in the
spin measurements is expected. All entangling Hamiltonians could
be, in principle, derived from fundamental principles, showing the
fundamental compatibility of quantum correlation measurements and
special relativity.

We thank H. Walther, C. Joachain, F. De Zela and N. Zagury for
useful comments.

\end{multicols}
\end{document}